\documentclass[11pt,a4paper]{article}
\usepackage{amsmath,amssymb,amsfonts,a4wide,graphicx,bm,times}
\usepackage{mcite}
\usepackage{tikz}

\usepackage{hyperref}
\hypersetup{
    colorlinks,
    citecolor=green,
    filecolor=green,
    linkcolor=blue,
    urlcolor=green
}

\makeatletter \let\old@startsection=\@startsection
\renewcommand{\@startsection}[6]
{\old@startsection{#1}{#2}{#3}{#4}{#5}{#6\mathversion{bold}}}
\makeatother

\makeatletter

\@addtoreset{equation}{section}
\makeatother
\let\refOld\ref
\renewcommand{\ref}[1]{(\refOld{#1})}

\newcommand{\superp}[2]{\genfrac{}{}{0pt}{}{#1}{#2}}
\def\res{\mathop{\text{Res}}}

 \def\d{\delta}
 
 \def\Re{{\rm Re ~}}
 
 \def\p{\partial}
 
 \def\a{\alpha}
 \def\b{\beta}
 \def\g{\gamma}
 \def\d{\delta}
 \def\e{\varepsilon}

 \def\l{\lambda}

 \def\s{\sigma}

 \def\z{\zeta }
 
 \def\G{\Gamma}

 \def\L{\Lambda}
 \def\O{\Omega}
 \def\o{\omega }
 
\def\CA{{\mathcal{A}}}
\def\CB{{\mathcal{B}}}

\def\CF{{\mathcal{F}}}

\def\CN{{\mathcal{N}}}
\def\CO{{\mathcal{O}}}

\def\CY{{\mathcal{Y}}}
\def\CZ{{\mathcal{Z}}}
\def\la{\left\langle}

\def\hf{\dfrac{1}{2}}

\def\sign{\text{sign }}

\def\qf{\mathfrak{q}}
\def\Zv{\mathcal{Z}_\text{vect.}}

\def\Zf{\mathcal{Z}_\text{fund.}}

\def\Zp{\mathcal{Z}_\text{pert.}}
\def\Zinst{\mathcal{Z}_\text{inst.}}
\def\CYY{\CY_{\vec Y}}

\def\rag{\right\rangle_\text{gauge}}

\def\aY{|\vec a,\vec Y\rangle}

\def\CFSW{\mathcal{F}_\text{SW}}

\begin{document}
\let\refOld\ref
\renewcommand{\ref}[1]{(\refOld{#1})}

\begin{titlepage}
\renewcommand{\thefootnote}{\fnsymbol{footnote}}
\vspace*{-2cm}
\begin{flushright}
KIAS-Q17053
\end{flushright}

\vspace*{1cm}

\begin{center}
{\huge {\bf  Seiberg-Witten period relations in Omega background}}

\vspace{10mm}
{\Large Jean-Emile Bourgine$^{\dagger \ast}$, Davide Fioravanti$^\ast$}
\\[.4cm]
{\em {}$^\dagger$Quantum Universe Center (QUC), KIAS}\\
{\em 85 Hoegiro, Dongdaemun-gu, Seoul, Republic of Korea}\\
[.4cm]
{\em {}$^\ast$ Sezione INFN di Bologna, Dipartimento di Fisica e Astronomia,
Universit\`a di Bologna} \\
{\em Via Irnerio 46, 40126 Bologna, Italy}\\
[.4cm]
\texttt{bourgine\,@\,kias.re.kr,\quad fioravanti\,@\,bo.infn.it}
\end{center}

\vspace{0.7cm}

\begin{abstract}
\noindent
Omega-deformation of the Seiberg-Witten curve is known to be written in terms of the qq-character, namely the trace of a specific operator acting in a Hilbert space spanned by certain Young diagrams. We define a differential form acting on this space and establish two discretised versions of the Seiberg-Witten expressions for the periods and related relations for the prepotential. 

\vspace{0.5cm}
\end{abstract}

\vfill

\end{titlepage}
\vfil\eject

\setcounter{footnote}{0}

\section{Introduction}
In 1994, Seiberg and Witten (SW) managed to characterise the IR behavior of $\CN=2$ 4d super Yang-Mills (SYM) theories \cite{Seiberg1994,Seiberg1994a}. In their work, a low energy effective Lagrangian is constructed from an holomorphic function of the superfields called the \textit{prepotential}, $\CFSW$. The prepotential itself is obtained from an hyper-elliptic curve of genus $N_c-1$, called after them \textit{Seiberg-Witten curve} and a specific differential form $dS_\text{SW}$ on it, the \textit{Seiberg-Witten differential} (whose singularities reveal the BPS spectrum of the theory). In details, the prepotential as an holomorphic function of the Coulomb branch parameters or periods, $a_l$, must generate, via Legendre transformation, the dual periods $a_{D,l}$ 
\begin{equation}\label{rel_SW}
a_l=\dfrac1{2i\pi}\oint_{\CA_l}dS_\text{SW},\quad 
a_{D,l}=\dfrac{\p\CFSW}{\p a_l}=\oint_{\CB_l}dS_\text{SW}
\end{equation} 
where $\CA_l$ and $\CB_l$ are two orthogonal sets of cycles on the SW curve ($l=1\cdots N_c$). Therefore, the prepotential can be regarded as the solution to the \textit{Seiberg-Witten relations} above. In this paper, we focus on $\CN=2$ super QCD theories, with a gauge group $U(N_c)$, and $N_f$ chiral matter multiplets in the fundamental representation. The explicit form of the SW curve and differential form for these theories will be given in next section \refOld{sec_NS}. The SW construction is closely related to classical integrable systems under the identification of the SW curve with the spectral curve of the latter \cite{Martinec:1995by, Donagi:1995cf}, of SW relations with the corresponding Whitham equations and the prepotential with the tau function ({\it cf.} \cite{Marshakov1999} and references therein). Besides, Whitham hierarchies describe the isomonodromic deformations of finite gap solutions (Hitchin systems) for integrable hierarchies of non-linear partial differential equations. In this way, the two SW relations play an essential role in the correspondence between the BPS sector of the gauge theory and classical integrable systems.

With the advance of localisation techniques, it has been possible to work out the exact expression of the partition function for $\CN=2$ SYM theories, including the non-perturbative instanton contributions \cite{Nekrasov2003,Losev2003}. In these computations, the theory is in the Euclidean space, but two IR regulators $\e_1$ and $\e_2$ are introduced to make finite the otherwise diverging volume of $\mathbb{R}^4$ while preserving supersymmetry. This particular background can be realised by turning on a graviphoton field, it is called the \textit{Omega background} (see \cite{Losev2003} for a detailed construction). When the two cut-offs are sent to zero, the logarithm of the partition function reproduces the SW prepotential, with a prefactor $(\e_1\e_2)^{-1}$ interpreted as the diverging volume of $\mathbb{R}^4$ \cite{Losev2003,Nekrasov2003}. In fact, the whole SW geometry, including the two relations \ref{rel_SW} can be recovered in this limit \cite{Nekrasov2003a}.

A lot of efforts have been devoted to extending the whole SW theory in presence of Omega background and to understanding its deep relationship with integrable systems. Mainly these efforts have been motivated by the study of the Nekrasov-Shatashvili (NS) limit of the background in which $\e_2$ is sent to zero while $\e_1$ remains finite. It has been observed that this limit of the gauge theory is described by a Thermodynamic Bethe Ansatz-like equation, {\it i.e.} a quantum integrable system at finite temperature/size \cite{Nekrasov2009}\footnote{We prove elsewhere \cite{Bourgine2017a} how this pseudoenergy function of the Thermodynamical Bethe Ansatz-like equation is actually connected to the counting function of a genuine non-linear integral equation \cite{Klumper1991,Destri, Fio} for the Bethe equations/integrable system, {\it cf.} also below the final section \refOld{conc}.}. The $\e_1$-deformation of the SW curve is identified with a sort of Baxter TQ-equation for the integrable system \cite{Poghossian2010,Fucito2011,Nekrasov2013}. Moreover, the $\e_1$-deformation of the differential form has been discussed in \cite{Bourgine2012,Bourgine2012a} using a perturbative approach (for small $\e_1$). The profound connection between the qSW geometry and quantum integrability has been a strong motivation for understanding further the role played by the $\e_2$-deformation.

Recently, new objects called \textit{qq-characters} have been introduced by Nekrasov in $\CN=2$ SYM theories \cite{Nekrasov2015,Nekrasov2016,Nekrasov2017} (see also \cite{Nekrasov2013,Bourgine2015c,Bourgine2016,Kim2016,Nekrasov2016a}). The condition that these quantities are polynomials in a spectral variable $z$ is indeed equivalent to the non-perturbative Schwinger-Dyson equations for a suitable resolvent. In addition, they possess an intriguing property that is directly relevant to the present discussion: they define a double deformation of the SW geometry \cite{Nekrasov2013,Bourgine2015c,Kimura2016}. Although their precise interpretation within the quantum integrability framework is not fully understood yet, they have an intricate relation to the underlying quantum algebras \cite{Bourgine2015c,Kimura2015,Bourgine2016,Kimura2016a}. In this paper, we focus on a different problem, that is to construct the double deformation of the SW differential form, and describe the SW relations at finite $\e_1$ and $\e_2$. One of the main difficulty lies in the ``discretization'' of the algebraic curve: for finite omega-background parameters $\e_1$ and $\e_2$, the branch cuts of the SW curve are replaced by series of poles, and the two sheets become disconnected. It is then necessary to define two differential forms to reconstruct the prepotential, related one to the other by some kind of symmetry.

In the next section, we shall provide some explicit expressions for the instanton partition function and the qq-character. In order to better describe the symmetry relating the two sheets of the curve, we shall briefly introduce the Spherical Hecke central algebra (SHc) acting on instanton partition functions, albeit this part is unnecessary at first reading. In section three, we shall tackle the problem of defining the qqSW differential form where we also infer the right qqSW relations involving the periods. Finally, we shall discuss the NS and SW limits of this doubly deformed geometry in section four, by recovering the  previously known results.

\section{Instanton partition function and qq-characters}\label{sec_Zinst}
\subsection{Instanton partition function}
In this paper, we focus on the 4d $\CN=2$ Super Yang-Mills with a $U(N_c)$ gauge group and $N_f$ matter fields in the fundamental (or anti-fundamental) representation with mass $m_f$, \,  $f=1,\dots, N_f$. The theory is considered on the Coulomb branch, and the vacuum expectation values (vevs) of the scalar field in the gauge multiplet will be denoted $a_l$ with $l=1\cdots N_c$. These parameters $a_l$ and $m_f$ are the roots of the gauge and matter polynomials respectively, namely
\begin{equation}
A(z)=\prod_{l=1}^{N_c}(z-a_l),\quad M(z)=\prod_{f=1}^{N_f}(z-m_f).
\end{equation}
We will assume $N_f<2N_c$ which entails an asymptotically free theory\footnote{The case $N_f=2N_c$, of asymptotically conformal theory, can be treated along the same lines, with only minor modifications.}. Then, the instanton contributions to the partition function in the Omega-background have been obtained in \cite{Nekrasov2003}. It takes the form of a sum over the $n$-instantons sectors of coupled integrals, that can be evaluated using Cauchy theorem as a sum of residues. The poles in the instanton sector $n$ are in one-to-one correspondence with the box configurations of $N_c$ Young diagrams $\vec Y=(Y^{(1)},\cdots, Y^{(N_c)})$ having a total number of boxes $|\vec Y|=n$. Precisely, the poles are located at the positions $\phi_x=a_l+(i-1)\e_1+(j-1)\e_2$ associated to the boxes $x=(l,i,j)$ of coordinates $(i,j)$ in the $l$th Young diagram $Y^{(l)}$. The corresponding residue can be factorized into contributions of gauge and matter multiplets, and the instanton partition function written formally as
\begin{equation}\label{Zinst_Ydiag}
\Zinst=\sum_{\vec Y}\qf^{|\vec Y|}\Zv(\vec a,\vec Y)\Zf(\vec m,\vec Y),
\end{equation} 
where $\vec Y$ are $N_c$-tuples Young diagrams, and $\vec a$, $\vec m$ vectors of dimension $N_c$ and $N_f$ respectively, encoding the dependence in Coulomb branch vevs and fundamental multiplets masses. The expression of the matter contribution to the residues is relatively simple, it is given as a product over all the boxes $x\in\vec Y$ of the mass polynomial evaluated at the poles location $\phi_x$. The usual expression for the vector multiplet contribution is fairly complicated, it can be found for instance in \cite{Bourgine2014a}. Here, we will use another expression that can be obtained by a direct evaluation of the residues, provided we introduce the UV regulator $\mu\in\mathbb{R}$ on the moduli space,\footnote{This expression can also be obtained using the formula derived in \cite{Bourgine2015c} by solving the discrete Ward identities,
\begin{equation}
\Zv(\vec a,\vec Y)=\prod_{x\in\vec Y}\left[A(\phi_x+\e_+)\CYY(\phi_x)\right]^{-1}.
\end{equation}
where the function $\CYY(z)$ is defined in \ref{def_CYY} below. In fact, this formula is well defined only if we express $\CYY(\phi_x)$ in terms of the shell ratio, i.e. the RHS of \ref{def_CYY}. In order to recover \ref{def_Zv}, we need to use the middle expression where the scattering factor $S(z)$ enters. Then, it is necessary to introduce the UV cut-off $\CYY(\phi_x)\to\CYY(\phi_x+\mu)$ to treat the factors in which instanton positions coincide up to $0,\pm\e_\a$ with $\a=1,2,+$.}
\begin{equation}\label{def_Zv}
\Zv(\vec a,\vec Y)=\lim_{\mu\to0}\prod_{x\in\vec Y}\left[A(\phi_x+\mu)A(\phi_x+\e_++\mu)\right]^{-1}\times\prod_{x,y\in\vec Y}S(\phi_x-\phi_y-\mu)^{-1}.
\end{equation} 
The UV regulator $\mu$ is reminiscent of the mass-dependence of the instanton contribution for matter fields in the adjoint representation \cite{Bourgine2015c}. The function $S(z)$ is sometimes called \textit{scattering factor}, it is defined as
\begin{equation}
S(z)=\dfrac{(z+\e_1)(z+\e_2)}{z(z+\e_+)}.
\end{equation} 



\subsection{Definition of the qq-character}
The qq-characters have been introduced in \cite{Nekrasov2015,Nekrasov2016,Nekrasov2017} as a generalization of the quantum group q-characters \cite{Knight1995,Frenkel1998,Frenkel2013}. They have the essential property to be polynomials of the spectral variable $z$. In fact, in analogy with matrix models they may also be interpreted as connected to the resolvent (somehow $\la\CYY(z)\rag$ below) satisfying the loop equations. The latter are called in this context non-perturbative Schwinger-Dyson equations and impose the qq-characters being a polynomial. In details, they can be defined as a trace over the instanton configurations of the following function depending on the detailed content of the Young diagrams
\begin{equation}\label{def_CYY}
\CYY(z)=A(z)\prod_{x\in\vec Y}S(\phi_x-z)=\dfrac{\prod_{x\in A(\vec Y)}(z-\phi_x)}{\prod_{x\in R(\vec Y)}(z-\e_+-\phi_x)}.
\end{equation} 
The second equality is the result of cancellations between factors associated to adjacent boxes. The RHS is expressed in terms of products over the sets $A(\vec Y)$ and $R(\vec Y)$ of boxes that can be added to or removed from the Young diagram vector $\vec Y$. Since these boxes are located on the edge of the Young diagrams, this formula is sometimes referred to as the \textit{shell formula}.

For the gauge theory of interest, the fundamental qq-character corresponds to
\begin{equation}\label{def_chi}
\chi(z)=\la\CYY(z+\e_+)+\qf \dfrac{M(z)}{\CYY(z)}\rag,
\end{equation} 
where the gauge average of any operator $X$ is defined via the measure of the sum over partitions as in the instanton partition function: 
\begin{equation}
\la \CO_{\vec Y} \rag=\dfrac1{\Zinst}\sum_{\vec Y}\qf^{|\vec Y|}\Zv(\vec a,\vec Y)\Zf(\vec m,\vec Y) \CO_{\vec Y} \ .
\end{equation} 

\subsection{Spherical Hecke central algebra}
Strictly speaking, the mention of the action of this affine Yangian algebra is not essential for establishing the qqSW relations. Yet, it provides an interesting insight on the role played by the two sheets of the SW curve in the discretized language. Readers not familiar with quantum algebras may be advised to skip this section at first reading.

The SHc algebra has been introduced by Shiffmann and Vasserot in \cite{Schiffmann2012}, it is known to act on the instanton partition function of $\CN=2$ 4d SYM \cite{Kanno2013}.\footnote{See also \cite{Prochazka2015} for the equivalence with the affine Yangian of $\mathfrak{gl}_1$.} It can be defined in terms of the generating series (see \cite{Bourgine2014a}),
\begin{equation}
D_{\pm1}(z)=\sum_{n=0}^\infty{z^{-n-1}D_{\pm1,n}},\quad D_0(z)=\sum_{n=0}^\infty{z^{-n-1} D_{0,n+1}},\quad E(z)=1+\e_+\sum_{n=0}^\infty{z^{-n-1} E_n},
\end{equation}
where $E(z)$ can be expressed using the modes of the current $D_0(z)$. This is done using the holomorphic operator $\Phi(z)$ as an intermediate step \cite{Bourgine2015c}:\footnote{Expressed in terms of modes $D_{0,n}$, we have
\begin{equation}
\Phi(z):=\log(z) D_{0,1}-\sum_{n=1}^\infty 
\frac{1}{n z^n}D_{0,n+1}.
\end{equation}}
\begin{equation}
D_0(z)=\p_z\Phi(z),\quad \CY(z)=A(z)e^{\Phi(z-\e_1)}e^{\Phi(z-\e_2)}e^{-\Phi(z)}e^{-\Phi(z-\e_+)},\quad E(z)=\CY(z+\e_+)\CY(z)^{-1},
\end{equation}
where the relation involves the Coulomb branch vevs $a_l$ considered as the weights of the representations. Since the modes $D_{0,n}$ are commuting, the inversion of the vertex operators $e^{\pm\Phi(z)}$ simply consists in flipping the sign in the exponential. Finally, the currents $D_{\pm1}$ and $D_0$ satisfy the following commutation relations,
\begin{equation}\label{comm_SHc}
[D_0(z),D_{\pm1}(w)]=\pm\dfrac{D_{\pm1}(w)-D_{\pm1}(z)}{z-w},\quad [D_{-1}(z),D_1(w)]=\dfrac{\e_1\e_2}{\e_+}\dfrac{E(w)-E(z)}{z-w}.
\end{equation}

The action of SHc on instanton partition functions is associated to the representation on a basis of states $\aY$ parameterized by $N_c$ Young diagrams $\vec Y$ in bijection with the instanton configurations arising in the expansion \ref{Zinst_Ydiag} of $\Zinst$. The action of SHc on these states can be written as follows,
\begin{align}
\begin{split}
&D_0(z)\aY=\sum_{x\in \vec{Y}}\dfrac1{z-\phi_x}\aY,\quad E(z)\aY=\dfrac{\CYY(z+\e_+)}{\CYY(z)}\aY,\\
&D_{1}(z)\aY=\sum_{x\in A(\vec{Y})}\dfrac{1}{z-\phi_x}\res_{z=\phi_x}\dfrac1{\CYY(z)}|\vec{a},\vec{Y}+x\rangle,\\
&D_{-1}(z)\aY=\sum_{x\in R(\vec{Y})}\dfrac{1}{z-\phi_x}\res_{z=\phi_x}\CYY(z+\e_+)|\vec{a},\vec{Y}-x\rangle,
\end{split}
\end{align}
where the function $\CYY(z)$, eigenvalue of the operator $\CY(z)$ on the state $\aY$, has been defined in \ref{def_CYY}. In this formalism, the instanton partition function is recovered as the norm of the Gaiotto state $|G,\vec{a}\rangle$ built as a coherent state in the SHc module.

The intimate connection between the qq-characters and the SHc algebra has been presented in \cite{Bourgine2015c}. To summarize, the regularity property of the qq-characters is a consequence of the particular transformation property of the Gaiotto state:
\begin{equation}
\label{action_Gaiotto}
D_{-1}(z)|G,\vec{a}\rangle=\dfrac1{\CY(z)}|G,\vec{a}\rangle,\quad D_{1}(z)|G,\vec{a}\rangle=\mathrm{P}^-_z\CY(z+\e_+)|G,\vec{a}\rangle
\end{equation}
where $\mathrm{P}_z^-$ denotes the projection on negative powers of $z$. The point we would like to stress here, is that the two terms in the expression \ref{def_chi} of the qq-character correspond to the action of either $D_1(z)$ or $D_{-1}(z)$. In the next section, we will define two different differential forms, using each of these two terms, and associate them to each of the two sheets of the Seiberg-Witten curve. 

\begin{figure}
\begin{center}
\begin{tikzpicture}
\draw[blue] (-0.6,0.6) -- (0,0) -- (1,0) -- (1.6,0.6);
\draw[blue] (-0.6,-1.6) -- (0,-1) -- (1,-1) -- (1.6,-1.6);
\draw[red] (0,-1) -- (0,0);
\draw[red] (1,-1) -- (1,0);
\end{tikzpicture}
\hspace{1cm}
\begin{tikzpicture}[scale=0.9]
\draw (-1.5,0) -- (1.5,0);
\draw (-1.5,-2) -- (1.5,-2);
\draw (0,0) -- (0,-2);
\node[right,scale=0.7] at (0,-1) {$(0,2)$};
\node[above,scale=0.7] at (-1.5,0) {$(1,n)$};
\node[above,scale=0.7] at (1.5,0) {$(1,n+2)$};
\node[below,scale=0.7] at (-1.5,-2) {$(1,n^\ast+2)$};
\node[below,scale=0.7] at (1.5,-2) {$(1,n^\ast)$};
\node[above,scale=0.7] at (0,0) {$\Phi$};
\node[below,scale=0.7] at (0,-2) {$\Phi^\ast$};
\end{tikzpicture}
\hspace{1cm}
\begin{tikzpicture}[scale=0.7]
\tikzset{>=latex}
\draw[<->] (0,1) -- (0,0) -- (1,0);
\node[right,scale=0.7] at (1,0) {$x^5$ (NS5)};
\node[above,scale=0.7] at (0,1) {$x^6$ (D5)};
\end{tikzpicture}
\end{center}
\caption{$(p,q)$-web diagram of 5d $\CN=1$ pure $U(2)$ gauge theory (left) and corresponding DIM representations web (right)}
\label{fig1}
\end{figure}
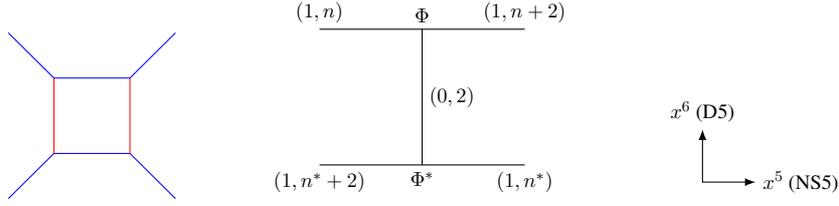

Algebraically, the exchange of the two sheets corresponds to exchanging the two generating series $D_{1}(z)$ and $D_{-1}(z)$. The algebra remains invariant provided that we flip the sign of the parameters $\e_1$ and $\e_2$ (and $D_0(z)$). This reflexion symmetry is the degenerate version of the symmetry $\s_V$ described in \cite{Bourgine2017d} for the Ding-Iohara-Miki (DIM) algebra \cite{Ding1997,Miki2007}. This algebra acts on the instanton partition functions of 5d $\CN=1$ super Yang-Mills theories, the uplift of 4d $\CN=2$ theories using an extra compact dimension. Correspondingly, the DIM algebra is the q-deformed version of SHc.\footnote{For more details on the relation between the DIM algebra and the qq-characters of 5d Super Yang-Mills theories, see \cite{Bourgine2016,Bourgine2017b}.} In this context, a dictionary has been established between the representations of DIM algebra and the $(p,q)$-brane engineering of the 5d gauge theories in type IIB string theory \cite{Mironov2016,Awata2016a,Bourgine2017b}. The symmetry $\s_V$ exchanging the DIM currents corresponds to a reflexion symmetry of the $(p,q)$-brane structure in the direction associated to the NS5-branes. In the case of a pure gauge theory ($N_f=0$), the 5d  $\CN=1$ uplift is obtained from two NS5 branes (bounded to D5-branes, in blue on figure \ref{fig1} left) linked by a stack of D5-branes (in red). The reflection $\s_V$ effectively exchanges the two NS5-branes, replacing the intertwiners (the algebraic version of the topological vertex) with their dual: $\Phi\leftrightarrow\Phi^\ast$.

A similar brane construction exists for 4d $\CN=2$ gauge theories, it exploits the type IIA string theory. In this construction, D5 and NS5 branes are replaced by D4 and solitonic 5 branes respectively. In this context, the two roots of the Seiberg-Witten curve equation are interpreted as the position of two solitonic 5-branes \cite{Witten1997}. Thus, the exchange of the two sheets of the SW curve corresponds to the reflection that exchanges the two five branes. This symmetry is the degenerate version of $\s_V$, which is indeed related to the exchange of the currents $D_1(z)$ and $D_{-1}(z)$, and by extension, to the exchange of the two terms in the definition of the qq-characters. 

%

\section{Derivation of the qq-Seiberg-Witten relations}
\subsection{qq-Seiberg-Witten differential forms and the first SW relation}
In order to generalise the SW formalism to the case with $\O$-backgrounds, we define a Young diagram valued differential form
\begin{equation}
dS_{\vec Y}=-\log(\CYY(z+\e_+))dz.
\end{equation} 
In the algebraic formalism described previously, this quantity would be naturally promoted to an operator $dS$ acting on the Hilbert space of states $\aY$, and diagonal in this basis with eigenvalue $dS_{\vec Y}$. In this definition, the argument of the logarithm coincides with the first term of the qq-character. In addition, we will associate to the second sheet of the SW curve, another differential form obtained from the second term of the qq-character,\footnote{The presence of $\qf$ in this definition is a purely aesthetic choice.}
\begin{equation}\label{def_SW}
dS_{\vec Y}^\ast=-\log\left(\dfrac{\qf M(z)}{\CYY(z)}\right)dz.
\end{equation}

Explicit expressions for these differential forms can be obtained by using the shell formula \ref{def_CYY}. These become sums over single poles after an integration by parts (represented here by the symbol $\equiv$):
\begin{align}
\begin{split}
&dS_{\vec Y}\equiv zdz\left(\sum_{x\in A(\vec Y)}\dfrac1{z+\e_+-\phi_x}-\sum_{x\in R(\vec Y)}\dfrac1{z-\phi_x}\right),\\
&dS_{\vec Y}^\ast\equiv -zdz\left(\sum_{x\in A(\vec Y)}\dfrac1{z-\phi_x}-\sum_{x\in R(\vec Y)}\dfrac1{z-\e_+-\phi_x}-\sum_{f=1}^{N_f}\dfrac1{z-m_f}\right).
\end{split}
\end{align}

With the exception of the poles at $z=m_f$ related to the presence of matter multiplets, all the poles lie inside the set $\{a_l+(i-1)\e_1+(j-1)\e_2\}$ for $(i,j)\in \mathbb{Z}^{\geq0}\times \mathbb{Z}^{\geq0}$ and $l=1\cdots N_c$. These discrete series of poles can be seen as the discretisation of the branch cuts in the principal sheet of the Seiberg-Witten curve \cite{Nekrasov2003a,Bourgine2012a}. In this perspective, it is natural to define for any $l$ the cycle $\CA_l$ that circles the subset $\{a_l+(i-1)\e_1+(j-1)\e_2\}$ at fixed $l$. A short calculation shows that 
\begin{equation}\label{qqSW_I}
\dfrac1{2i\pi}\la\oint_{\CA_l}dS_{\vec Y}\rag=\la\sum_{x\in A(Y^{(l)})}(\phi_x-\e_+)-\sum_{x\in R(Y^{(l)})}\phi_x\rag=a_l-\e_+.
\end{equation}
This equality can be interpreted as the first of the two qq-Seiberg-Witten relations. Similarly, on the second sheet, we can write down
\begin{equation}
\dfrac1{2i\pi}\la\oint_{\CA_l}dS_{\vec Y}^\ast\rag=-a_l.
\end{equation}

\subsection{The second qqSW relation}
The second SW relation involves the gauge theory prepotential, defined in the Omega-background as the logarithm of the partition function,
\begin{equation}
\CF=-\e_1\e_2\log\CZ,\quad \CZ=\Zp\Zinst.
\end{equation} 
The partition function entering this definition is the product of a perturbative factor $\Zp$ and the instanton corrections $\Zinst$ discussed in the section \refOld{sec_Zinst}. For simplicity we restrict ourselves to the case of the gauge group $SU(N_c)$, that corresponds to impose the vanishing of the sum of the Coulomb branch vevs, $\sum_la_l=0$. Due to the presence of $\CN=2$ supersymmetry, the perturbative contribution is one-loop exact, and the derivation of the one loop determinant can be found in \cite{Losev2003}. The vector multiplet contribution takes the form of a quadruple infinite product. Its expression is simplified by the introduction of the notation $\vec Y_\infty$ that consists in $N_c$ Young diagrams of infinite size: $\vec Y_\infty=\{(l,i,j)\}$ with $l=1\cdots N_c$ and $i,j\in\mathbb{Z}^{>0}$. The matter multiplet contribution can be found in \cite{Alday2009} expressed in terms of the Barnes double gamma function, which is simply the regularization of the double infinite product over the boxes in $\vec Y_\infty$. Hence, the perturbative contribution can be written formally as
\begin{equation}\label{def_Zp}
\Zp=\qf^{-\frac1{2\e_1\e_2}\left(\sum_la_l^2-\frac{N_c}{2N_c-N_f}\sum_fm_f(m_f+\e_+)\right)}\prod_{x,y\in\vec Y_\infty}S(\phi_x-\phi_y)^{-1}\times\prod_{x\in\vec Y_\infty}M(\phi_x)^{-1}.
\end{equation} 
This formal expression requires two types of regularization. As in the case of the instanton contribution, a UV cut-off on the moduli space is required to remove the singularities when two instantons approach at a distance $d\sim 0,\e_\a$ in the product of scattering factors $S(\phi_x-\phi_y)$. In the appendix \refOld{AppA}, a deformation of the scattering factor into $S(\phi_x-\phi_y-\mu)$ is introduced, and it is shown that the infinite product multiplied by the factor $\mu^{N_c}$ remains finite in the limit $\mu\to0$. In addition, an IR regularization is also required to remove the singularities coming from the infinite size of the Young diagrams in $\vec Y_\infty$. This is done using a $\z_2$-regularization, so that we recover the usual expression in terms of the logarithm of Barnes double gamma functions \cite{Losev2003},
\begin{equation}
\Zp=\prod_{l,l'=1}^{N_c}
e^{-\g_{\e_1,\e_2}(a_l-a_{l'}-\e_+|\L)}\times\prod_{l=1}^{N_c}\prod_{f=1}^{N_f}e^{\g_{\e_1,\e_2}(a_l-m_f-\e_+|\L)}
\end{equation}
with $\qf=\L^{2N_c-N_f}$, and up to an irrelevant, $(a_l,m_f)$-independent, constant factor. A brief reminder on the function $\g_{\e_1,\e_2}(z|\L)$ is given in appendix \refOld{AppA}. However, in the following we will mostly work with the formal expression \ref{def_Zp}, keeping in mind both IR and UV regularizations.

The expression of the full partition function is the product of the perturbative part given in \ref{def_Zp} and the instanton contributions written in \ref{Zinst_Ydiag}. Simplification occurs if we write the gauge polynomial $A(z)$ entering the vector multiplet factor \ref{def_Zv} in the form of an infinite product using the formula \ref{prop_A_G2} given in appendix,
\begin{equation}
\Zv(\vec a,\vec Y)=\prod_{\superp{x\in\vec Y}{y\in\vec Y_\infty}}S(\phi_x-\phi_y)\times \prod_{\superp{x\in\vec Y_\infty}{y\in\vec Y}}S(\phi_x-\phi_y)\times\prod_{\superp{x\in\vec Y}{y\in\vec Y}}S(\phi_x-\phi_y)^{-1},
\end{equation} 
where IR and UV regularizations have been omitted. As a result, the partition function $\CZ$ can be written as products over the infinite sets of boxes $\vec Y_c$ that are complementary to the Young diagrams $\vec Y$ in the infinite lattice $\vec Y_\infty$,
\begin{equation}\label{def_Z}
\CZ=\qf^{-\frac1{2\e_1\e_2}\left(\sum_la_l^2-\frac{N_c}{2N_c-N_f}\sum_fm_f(m_f+\e_+)\right)}\sum_{\vec Y}\qf^{|\vec Y|}\prod_{x,y\in\vec Y_c}S(\phi_x-\phi_y)^{-1}\times\prod_{x\in\vec Y_c}M(\phi_x)^{-1},
\end{equation} 
with $\vec Y_c=\vec Y_\infty\setminus\vec Y$. In this form, the resemblance between the full partition function and its perturbative part is striking: the products involving the infinite Young diagrams $\vec Y_\infty$ in \ref{def_Zp} are replaced by a summation over all the possible finite sub-diagrams $\vec Y$ of products over the truncations $\vec Y_c=\vec Y_\infty\setminus\vec Y$.\footnote{The phenomenon resemble a melting crystal, with instanton corrections progressively removing boxes from the corner of an infinite stack of boxes, effectively melting the infinite partition.} A similar idea is already present in the work of Nekrasov and Okounkov \cite{Nekrasov2003a}, where the effective action describing the full partition function involves the same function $\g_{\e_1,\e_2}(z|\Lambda)$  than the perturbative contribution. Here, the role devoted to the Young diagram profiles is played by products over the boxes in $\vec Y_c$, and the special function $\g_{\e_1,\e_2}(z|\Lambda)$ is replaced by the logarithm of the scattering function $\log S(z)$ using its relation with the double gamma function and the functional relation \ref{prop_G2}. This idea has been further discussed in the context of the Nekrasov-Shatashvili limit in \cite{Bourgine2012a}.

Before stating the derivation of the qqSW relation, we would like to make a comment about the dependence in the gauge coupling $\qf$. In asymptotically free theories where $N_f<2N_c$, the partition function enjoys a scaling invariance under the transformation of the parameters $a_l\to \L a_l$, $m_f\to \L m_f$, $\e_\a\to\L\e_\a$ (and $\phi_x\to\L\phi_x$) and $\qf\to \L^{-(2N_c-N_f)}\qf$. This is a consequence of the transformation properties $\g_{\e_1,\e_2}(z|\L)=\g_{\e_1/\L,\e_2/\L}(z/\L|1)$ for the perturbative part, and
\begin{equation}
\Zv(\vec a,\vec Y)\to \L^{-2N_c|\vec Y|}\Zv(\vec a,\vec Y),\quad \Zf(\vec m,\vec Y)\to \L^{N_f|\vec Y|}
\end{equation} 
for the instanton part. Choosing the rescaling factor such that $\qf=\L^{2N_c-N_f}$, the dependence in the gauge coupling $\qf$ disappears. From now on, we will assume that this re-scaling has been done and set $\qf=1$.

From the expression \ref{def_Z}, it is easy to compute the derivative of the prepotential with respect to the Coulomb branch vev $a_l$,
\begin{equation}
\dfrac{\p\CF}{\p a_l}=-\e_1\e_2\la\sum_{\superp{x\in Y_c^{(l)}}{y\in \vec Y_c}}\left[\s(\phi_y-\phi_x)+\s(\phi_y-\phi_x-\e_+)\right]-\sum_{x\in Y_c^{(l)}}\s_m(\phi_x)\rag,
\end{equation} 
where $\s(z)=-\s(-z-\e_+)$ is the logarithmic derivative of the scattering function and $\s_m(z)$ the logarithmic derivative of the matter polynomial,
\begin{equation}
\s(z)=\p_z\log S(z),\quad \s_m(z)=\p_z\log M(z).
\end{equation} 
As in the expression of the partition function, the infinite sums are defined using $\z_2$-regularization, and the singularities at short distances can be removed using the UV cut-off $\mu$.

The next step is to relate the derivative of the prepotential to the qq-Seiberg-Witten differential. In \cite{Nekrasov2003a}, a saddle point technique is employed to perform the $\e_1,\e_2\to0$ limit, and relate the saddle point equation to the Seiberg-Witten curve. In contrast, there is no saddle point in the case of finite $\e_1,\e_2$, and the full average over partitions $\vec Y$ must be kept while the Seiberg-Witten differential is promoted to an operator. This operator has the eigenvalues $dS_{\vec Y}$ given in \ref{def_SW} and expressed in terms of the function $\CYY(z)$. The logarithmic derivative of this function can be computed using the definition \ref{def_CYY} and the expression \ref{prop_A_G2} of the gauge polynomial,
\begin{equation}
\dfrac{\p}{\p z}\log\CYY(z)=\sum_{x\in \vec Y_c}\s(\phi_x-z).
\end{equation}
This result allows us to express the $a_l$-derivative of the prepotential in terms of the derivative of the two differential forms:
\begin{equation}\label{qqSW_II}
\dfrac{\p\CF}{\p a_l}=\e_1\e_2\la\sum_{x\in Y_c^{(l)}}\left[\left.\dfrac{d^2S_{\vec Y}(z)}{dz^2}\right|_{z=\phi_x}-\left.\dfrac{d^2S_{\vec Y}^\ast(z)}{dz^2}\right|_{z=\phi_x}\right]\rag.
\end{equation}

The expression \ref{qqSW_II} should be compared with the second Seiberg-Witten relation \ref{rel_SW}, in which the cycle $\CB_l$ starts on the second sheet at $\infty_-$, goes to an extremity of the $l$th branch cut, then through the branch cut on the first sheet, and finally to $\infty_+$. At finite $\e_1,\e_2$, the contour integral from infinity to the branch cut is replaced by the discrete sum over the set $Y_c^{(l)}$. In order to ease the comparison with the contour integral, we decompose $Y_c^{(l)}=Y_\infty^{(l)}\setminus Y^{(l)}$, and write
\begin{equation}\label{qqSW_III}
\dfrac{\p\CF}{\p a_l}=\e_1\e_2\la \sum_{x\in Y_\infty^{(l)}}\left.\dfrac{d^2S_{\vec Y}(z)}{dz^2}\right|_{z=\phi_x}-\sum_{x\in Y^{(l)}}\left.\dfrac{d^2S_{\vec Y}(z)}{dz^2}\right|_{z=\phi_x}+ \sum_{x\in Y^{(l)}}\left.\dfrac{d^2S_{\vec Y}^\ast(z)}{dz^2}\right|_{z=\phi_x}-\sum_{x\in Y_\infty^{(l)}}\left.\dfrac{d^2S_{\vec Y}^\ast(z)}{dz^2}\right|_{z=\phi_x}
\rag.
\end{equation}
The first two terms are naturally associated to the part of the contour lying on the first sheet. The third and the last terms correspond to the part of the contour on the second sheet, they are indeed expressed using the differential form $dS^\ast$.

We have seen that, at finite $\e_1,\e_2$, the contour integral of the differential form $dS$ over $\CB_l$ is discretized into a sum over boxes $x$ of the double derivative $d^2S/dz^2$ evaluated at $z=\phi_x$. In fact, there are two successive discretizations, each associated to one parameter $\e_\a$, which explains the presence of a double derivative, and the pre-factor $\e_1\e_2$. In the next section, we discuss the limit $\e_1,\e_2\to0$ and show how the differential form \ref{qqSW_III} reduces to the original definition of Seiberg and Witten.

\paragraph{Relation with SHc generators:} We would like to conclude this section with a brief remark on the connection with the representation of the SHc algebra. Working in a single plane, the derivative of the prepotential is expressed using the difference $dS^-=dS-dS^\ast$ of the differential forms. On the other hand, the sum $dS^+=dS+dS^\ast$ is related to the generator $E(z)$ in the Cartan of SHc,
\begin{equation}
dS^+=-\log(\qf M(z)E(z))dz,
\end{equation} 
where the logarithm of a diagonal operator is defined by decomposition on its eigenvalues in the basis $\aY$. Using an integration by parts, the differential forms can also be written in terms of $D_0(z)$, for instance
\begin{equation}
dS\equiv z\left(\sum_{l=1}^{N_c}\dfrac1{z-a_l}-(1-e^{\e_1\p_z})(1-e^{\e_2\p_z})D_0(z)\right)dz.
\end{equation} 


\section{Nekrasov-Shatashvili and Seiberg-Witten limiting theories}\label{sec_NS}

\subsection{The Nekrasov-Shatashvili limit}
The NS limit of the theory is obtained by sending the Omega background parameter $\e_2$ to zero. There exists two different limiting procedures. The first approach takes the limit of the integral expression in \cite{Nekrasov2003} using Mayer cluster expansion \cite{Nekrasov2009,Meneghelli2013,Bourgine2014}.\footnote{This procedure has been refined in \cite{Bourgine2015,Bourgine2015a} in order to derive the subleading corrections in $\e_2$.} The second one, developed in \cite{Poghossian2010,Fucito2011,Fucito2012,Nekrasov2013,Bourgine2014a}, is based on a discrete saddle point technique applied to the Young diagram expansion \ref{Zinst_Ydiag}. This is the approach relevant to the discussions presented in this section. The relation between the two different approaches will be discussed in the companion paper \cite{Bourgine2017a}.

In this limit, the instanton partition function is described by a system of Bethe equations in the thermodynamical limit. In the limit $\e_2\to0$, the qq-character $\chi(z)$ remains polynomial. It plays the role of the Baxter T polynomial, and it will be denoted $h(z)$. Furthermore, the vev of the $\CYY$-operator can be expressed as a ratio of the function $v(z)$ that defines a regularized Baxter Q-polynomial in the thermodynamical limit:
\begin{equation}
\la\CYY(z)\rag\to Y(z)=\dfrac{v(z)A(z)}{v(z-\e_1)}.
\end{equation} 
The functions $h(z)$ and $v(z)$ obey a sort of TQ-equation, called $hv$-equation in \cite{Bourgine2017a},
\begin{equation}\label{hv}
\dfrac{h(z)}{A(z+\e_1)}v(z)=v(z+\e_1)+\qf Q(z)v(z-\e_1).
\end{equation}
This equation is obtained from the NS limit of the definition \ref{def_chi} of the qq-character. It is interpreted as the first quantization of the Seiberg-Witten curve, with $\e_1$ playing the role of the Planck constant \cite{Poghossian2010,Fucito2011}.

The NS limit is obtained from a discrete saddle point approximation. As a result, the gauge vev of the logarithm of the operator $\CY(z)$ becomes equal to the logarithm of the vev. From this result, it is easy to perform the limit of the vev of the qqSW differential forms. In particular, the difference $dS^-=dS-dS^\ast$ reproduces the qSW differential defined in \cite{Bourgine2012a} from an $\e_1$-perturbative approach around the Seiberg-Witten solution.\footnote{In \cite{Bourgine2012a}, the notation $\o(z)=Y(z)^{-1}$ has been employed.},
\begin{equation}
\la dS_{\vec Y}^-\rag \xrightarrow{\e_2\to0}dS_\text{NS}^-=\log\left[\dfrac{\qf M(z)}{Y(z)Y(z+\e_1)}\right]dz,
\end{equation}
It was shown in this paper that this differential form leads to q-deformed SW relations.

In order to take the limit of the qqSW relations, we need to provide more details on the behavior of the gauge vev of operators as $\e_2\to0$. In this limit, the sum over Young diagrams $\vec Y$ is dominated by the weight of a particular $N_c$-tuple diagram $\vec Y^\ast$ of infinite size. More precisely, the number of boxes $\l_i^{(l)}$ in each column is infinite, while the product $\e_2\l_i^{(l)}$ remains finite. Thus, the variables $j$ associated to the position of the box $x=(l,i,j)$ becomes continuous. On the other hand, the variable $i$ remains discrete, although the number of columns is also infinite. In this limit, the shape of the Young diagrams is encoded in the variables $u_{(l,i)}=a_l+(i-1)\e_1+\e_2\l_i^{(l)}$ identified with the Bethe roots of the integrable system \cite{Bourgine2017a}. As a result, we expect that the second qSW relation takes the form
\begin{equation}\label{limit_qqSW_I}
\e_1\e_2\la\sum_{x\in Y_c^{(l)}}\left.\dfrac{d^2S_{\vec Y}^-(z)}{dz^2}\right|_{z=\phi_x}\rag\xrightarrow{\e_2\to0}\e_1\sum_{i=1}^\infty\int_{u_{(l,i)}}^\infty\dfrac{d^2S_\text{NS}^-(t)}{dt^2}dt.
\end{equation} 
So far, the qSW relation has not been established in this context, and we hope to come back to this problem in a future publication.

\subsection{To the Seiberg-Witten theory}\label{sec42}
Taking further the limit $\e_1\to0$, we ought to reproduce SW theory:
\begin{equation}
\chi(z)\xrightarrow{\e_2\to0}h(z)\xrightarrow{\e_1\to0}P(z)=A(z)+O(\qf),\quad \la\CYY(z)\rag\xrightarrow{\e_2\to0} Y(z)\xrightarrow{\e_1\to0} y(z).
\end{equation} 
Note that the qq-character reduces to the polynomial $P(z)$ that correspond to the vacuum expectation value of the characteristic polynomial associated to the Higgs field, namely the scalar field of the vector multiplet \cite{Nekrasov-Pestun}. It differs from the gauge polynomial $A(z)$ by instanton corrections. The $hv$-equation \ref{hv} can be written in the form
\begin{equation}
h(z)=Y(z+\e_1)+\qf\dfrac{M(z)}{Y(z)},
\end{equation} 
it reproduces the SW curve in the SW limit, writing $y=\sqrt{M}w$:
\begin{equation}
\dfrac{P(z)}{\sqrt{M(z)}}=w(z)+\dfrac{\qf}{w(z)}.
\end{equation}
The two solutions of this quadratic equation are given by
\begin{equation}\label{sol_SW_curve}
w_\pm(z)=\dfrac{P(z)}{2\sqrt{M(z)}}\left(1\pm\sqrt{1-4\qf\dfrac{M(z)}{P(z)^2}}\right),
\end{equation}
they satisfy the relation $w_+w_-=\qf$. Correspondingly, writing $y_\pm=\sqrt{M}w_\pm$ we have $y_+y_-=\qf M$. Perturbatively in $\qf$, the two solutions behave differently:
\begin{equation}
y_+(z)=A(z)+O(\qf),\quad y_-(z)=\dfrac{\qf M(z)}{A(z)}+O(\qf^2).
\end{equation} 
In the NS limit, the function $Y(z)$ can be computed perturbatively in $\qf$, it behaves as $Y(z)=A(z)+O(\qf)$ (see \cite{Bourgine2017a}). Thus, in the limit $\e_1\to0$, the function $y(z)$ has to be identified with the solution $y_+(z)$ of the SW curve equation. As a result, the limit of the vev of the differential form $dS$ reproduces the SW form on the first sheet. We also observe that in this limit, the vev of the differential form $dS^\ast$ coincides with the expression of the SW form on the second sheet:
\begin{align}
\begin{split}
&\la dS_{\vec Y}\rag\xrightarrow{\e_1,\e_2\to0} -\log(y_+(z))dz\equiv z\dfrac{dy_+}{y_+},\\
&\la dS^\ast_{\vec Y}\rag\xrightarrow{\e_1,\e_2\to0} -\log\left(\dfrac{\qf M(z)}{y_+(z)}\right)dz=-\log(y_-(z))dz\equiv z\dfrac{dy_-}{y_-}.
\end{split}
\end{align}

Finally, it remains to prove that we recover the proper cycles for the contour integrals. The cycle $\CA_l$ has been defined as the cycle surrounding the poles located at $z=\phi_x$ for $x\in Y_\infty^{(l)}$. In the NS limit, it is sufficient to consider the poles associated to the Young diagrams $\vec Y^\ast$ that dominate the saddle point. In fact, the contour can be restricted to the singularities of the SW differential, i.e. $z=\phi_x-\e_+$ for $x\in A(Y^{\ast(l)})$ and $z=\phi_x$ for $x\in R(Y^{\ast(l)})$. The corresponding density reproduces the profile of the partition defined in \cite{Nekrasov2003a} (up to a shift of $\e_+$ in the spectral parameter, irrelevant in the SW limit):\footnote{As a consequence, the qqSW differential form is related to the resolvent associated to the density $\sum_l f''_l(t)$:
\begin{equation}
\dfrac{d^2S_{\vec Y}}{dz^2}=-\hf\int{\dfrac{f''(t)}{z+\e_+-t}dt}.
\end{equation}}
\begin{align}
\begin{split}\label{def_profile}
f_l(t)&=|t-a_l|-(1-e^{-\e_1\p_t})(1-e^{-\e_2\p_t})\sum_{x\in Y^{(l)}}|t-\phi_x|\\
&=\sum_{x\in A(Y^{(l)})}|t-\phi_x|-\sum_{x\in R(Y^{(l)})}|t-\e_+-\phi_x|.
\end{split}
\end{align}
In the SW limit, the $l$th branch cut of the SW curve coincides with the support of the second derivative of the profile of the Young diagram $Y^{\ast(l)}$. This branch cut is the content of the cycle $\CA_l$.

On the other hand, the SW limit of the second qqSW relation is technically a little more involved. It is given in the appendix \refOld{AppB}.

\section{Conclusions and discussions}\label{conc}
We have proven that the prepotential of $\CN=2$ super Yang-Mills on Omega background obeys the (doubly deformed SW) relations
\begin{equation}
a_l-\e_+=\dfrac1{2i\pi}\la\oint_{\CA_l}dS_{\vec Y}\rag,\quad \dfrac{\p\CF}{\p a_l}=\e_1\e_2\la\sum_{x\in Y_c^{(l)}}\left(\left.\dfrac{d^2S_{\vec Y}(z)}{dz^2}\right|_{z=\phi_x}-\left.\dfrac{d^2S_{\vec Y}^\ast(z)}{dz^2}\right|_{z=\phi_x}\right)\rag.
\end{equation} 
In the second relation, the prepotential is written in terms of the difference $dS-dS^\ast$ which is essentially given by the logarithm of the product $\CY(z)\CY(z+\e_+)$ of the operator $\CY(z)$ entering in the definition of the qq-character \ref{def_chi}. This operator also plays an important role in the study of Frenkel-Reshetikhin q-characters \cite{Frenkel1998}. However, it is the ratio $\CY(z+\e_+)/\CY(z)$ that relates itself to the Cartan generator of the Spherical Hecke central algebra. Even so, these newly defined operator $dS$ and $dS^\ast$ are expected to play an important role in the correspondence with quantum integrability which still deserves further study.

In \cite{Eynard2017} a new formalism has been developed, in which the Seiberg-Witten relations appear as an example of invariance under deformations associated to the cycles of a spectral curve. The qq-Seiberg-Witten relation presented here may provide a hint on the double quantization of this formalism.

Interestingly, in the NS limit, the qqSW differential has a simple expression in terms of the so-called counting function of the non-linear integral equation \cite{Destri, Fio} for the corresponding integrable system, $dS_\text{NS}=i\pi\eta(z)dz$ \cite{Bourgine2017a}  \footnote{As this reference proves, the counting function is linked to the pseudoenergy of the initial Thermodynamical Bethe Ansatz-like equation in \cite{Nekrasov2009}.}. Using the identification of the prepotential with the Yang-Yang functional, we should be able to derive the qSW relations in this context. Taking the limit $\e_2\to0$ of our results, we expect a relation of the form
\begin{equation}
\dfrac{\p\CF}{\p a_l}=2i\pi\e_1\sum_{i=1}^{\infty}\int_{u_{(l,i)}}^\infty \eta'(z)dz.
\end{equation}
We shall be able to give a detailed interpretation of the qSW relations applied to quantum integrable systems in a forthcoming publication.

The generalisation of our results to quiver gauge theories should be rather straightforward. A more challenging objective would be to derive these relations for $\CN=2$ theories with SO/Sp gauge groups. The main difficulty comes from the absence of a weak coupling expansion over Young diagrams configurations. We hope to address this issue in the near future.

\section*{Acknowledgments}

We would like to thank F. Bastianelli, A. Bonini, A. Fachechi, M. Rossi and A. Sciarappa for discussions. This project was partially supported by the grants: (I.N.F.N. IS) GAST (which supported JEB by an I.N.F.N. post-doctoral fellowship), UniTo-SanPaolo Nr TO-Call3-2012-0088, the ESF Network HoloGrav (09-RNP-092 (PESC)), MPNS--COST Action MP1210 and the EC Network Gatis.

\appendix

\section{IR/UV regularization of infinite products}\label{AppA}
\subsection{Barnes double gamma function}
Our conventions and notations for these special functions will be borrowed from \cite{Spreafico}, but we will omit the dependence in the two pseudo-periods $\e_1$ and $\e_2$. The Barnes double zeta function $\z_2(s|z)$ is defined as a double sum for $\Re(s)>2$, but it can be analytically continued using an integral expression obtained by Mellin transform,
\begin{equation}
\z_2(s|z)=\sum_{i,j=1}^\infty(z+(i-1)\e_1+(j-1)\e_2)^{-s}=\dfrac1{\G(s)}\int_0^\infty{\dfrac{dt}{t}t^s\dfrac{e^{-tz}}{(1-e^{-\e_1 t})(1-e^{-\e_2 t})}}.
\end{equation}
The integral in the RHS is well-defined for $\e_1$ and $\e_2$ real positive and it must be properly analytically continued for general values of the Omega-background parameters.

In \cite{Nekrasov2003a}, the perturbative contribution has been expressed using the function 
\begin{equation}
\g_{\e_1,\e_2}(z|\L)=\left.\dfrac{d}{ds}\right|_{s=0}\dfrac{\L^s}{\G(z)}\int_0^\infty{\dfrac{dt}{t}t^s\dfrac{e^{-tz}}{(e^{\e_1 t}-1)(e^{\e_2 t}-1)}}.
\end{equation} 
The dependence in $\L\in\mathbb{R}^+$ can be absorbed in a change of variable for $t$, that leads to the scaling property $\g_{\e_1,\e_2}(z|\L)=\g_{\e_1/\L,\e_2/\L}(z/\L|1)$. It can also be computed from Leibniz's product rule using the value of $\z_2(0|z)$ given in \cite{Spreafico},
\begin{equation}
\g_{\e_1,\e_2}(z|\L)=\left(\dfrac{(z-\e_1)(z-\e_2)}{2\e_1\e_2}+\dfrac{\e_+^2}{12\e_1\e_2}-\dfrac1{12}\right)\log\L+\g_{\e_1,\e_2}(z|1).
\end{equation} 
The function $\g_{\e_1,\e_2}(z|1)$ coincides with the $s$-derivative of the function $\z_2(s|z)$ evaluated at the point $s=0$ (up to a shift in the argument), and it corresponds to the logarithm of Barnes $\G_2(z)$ function up to a normalization factor,
\begin{equation}
\z_2'(0|z)=\g_{\e_1,\e_2}(z-\e_+|1)=\log\G_2(z)+\chi_2'(0),
\end{equation} 
where $\chi_2'(0)$ is identified with the derivative of the Riemann-Barnes double zeta function using the Lerch formula \cite{Spreafico},
\begin{equation}
\chi_2(s)=\sum_{\superp{i,j=0}{(i,j)\neq(0,0)}}^\infty (i\e_1+j\e_2)^{-s}.
\end{equation}

The standard gamma function is well-known to provide a regularization of infinite products through the Weierstrass definition
\begin{equation}
\G(z)=\dfrac{e^{-\g z}}{z}\prod_{n=1}^\infty e^{z/n}\dfrac{n}{z+n}.
\end{equation} 
Similarly, the double gamma function can be used to regularize infinite double products,
\begin{equation}\label{G2_product}
\G_2(z)=\dfrac1{z}e^{-r_1 z+\frac12 r_2 z^2}\prod_{\superp{i,j=0}{(i,j)\neq (0,0)}}^{\infty}e^{\frac{z}{i\e_1+j\e_2}-\frac{z^2}{2(i\e_1+j\e_2)^2}}\dfrac{i\e_1+j\e_2}{z+i\e_1+j\e_2},
\end{equation} 
where the coefficients $r_1$ and $r_2$ are related to the residues of the function $\chi_2(s)$ at $s=1$ and $s=2$ respectively. Their explicit expression, fairly complicated, will not be given here, but it can be found in \cite{Spreafico}. This extension (or $\b$-deformation) of the gamma function to a second pseudo-period leads to a functional relation slightly more involved,
\begin{equation}\label{prop_G2}
\dfrac{\G(z+1)}{\G(z)}=z\quad\to\quad \dfrac{\G_2(z+\e_1)\G_2(z+\e_2)}{\G_2(z)\G_2(z+\e_+)}=z.
\end{equation} 

\subsection{UV cut-off}
In order to discuss the UV cut-off, we introduce another large size regularization for the infinite Young diagrams: we simply truncate them to be squares of size $N\times N$. So, let $\vec Y_N$ be the collection of $N_c$ Young diagrams filled with $N$ columns of $N$ boxes, we have
\begin{equation}
A(\vec Y_N)=\{(l,N+1,1),(l,1,N+1)\},\quad R(\vec Y_N)=\{(l,N,N)\},\quad l=1\cdots N_c,
\end{equation} 
and the shell formula gives
\begin{equation}\label{shell_YN}
\prod_{x\in\vec Y_N}S(\phi_x-z)=\dfrac{A(z-N\e_1)A(z-N\e_2)}{A(z)A(z-N\e_+)}.
\end{equation}

This expression can be used to derive a representation of the gauge polynomial as an infinite product. Indeed, sending the IR cut-off $N$ to infinity, we find that
\begin{equation}
\prod_{x\in\vec Y_N}S(\phi_x-z)\sim\left(-\dfrac{\e_1\e_2}{\e_+}N\right)^{N_c}\dfrac1{A(z)}.
\end{equation}
In fact, this representation can also be derived using the $\z_2$-regularization, combining the property \ref{prop_G2} of the $\G_2$-function and its representation \ref{G2_product} as infinite products,
\begin{equation}\label{prop_A_G2}
A(z)=(-1)^{N_c}\prod_{l=1}^{N_c}\dfrac{\G_2(a_l-z+\e_1)\G_2(a_l-z+\e_2)}{\G_2(a_l-z)\G_2(a_l-z+\e_+)}=(-1)^{N_c}\prod_{x\in\vec Y_\infty}S(\phi_x-z)^{-1}.
\end{equation} 
where the product in the RHS is $\z_2$-regularized.

To discuss the nature of UV-singularities, we come back to the shell formula \ref{shell_YN} applied to the Young diagrams $\vec Y_N$, and consider the double product
\begin{equation}
\prod_{x,y\in\vec Y_N}S(\phi_{xy}-\mu)=\prod_{l,l'=1}^{N_c}\prod_{i,j=1}^{N}\dfrac{(a_{ll'}+\mu-N\e_1+(i-1)\e_1+(j-1)\e_2)(a_{ll'}+\mu-N\e_2+(i-1)\e_1+(j-1)\e_2)}{(a_{ll'}+\mu+(i-1)\e_1+(j-1)\e_2)(a_{ll'}+\mu-N\e_++(i-1)\e_1+(j-1)\e_2)},
\end{equation}
where we have denoted $a_{ll'}=a_l-a_{l'}$. The only singular terms at $\mu=0$ have the indices $l=l'$ and $i=j=1$, so that the limit $\mu\to0$ of 
\begin{equation}
\mu^{N_c}\prod_{x,y\in\vec Y_N}S(\phi_{xy}-\mu)
\end{equation}
is finite. The exponent of $\mu$ is independent of the IR cut-off $N$, the two limits $N\to\infty$ and $\mu\to0$ are independent and we expect that
\begin{equation}
\mu^{N_c}\prod_{x,y\in\vec Y_\infty}S(\phi_{xy}-\mu)
\end{equation}
has a finite limit as well when $\mu$ is sent to zero.

\section{Limit $\e_1,\e_2\to0$ of the second Seiberg-Witten relation}\label{AppB}
In this section, we will follow the assumptions made in \cite{Nekrasov2003a} and consider $\e_1$ and $\e_2$ real, so that the branch cuts of the SW curve lie on the real line.The SW differentials on the two sheets will be denoted by $dS_\text{SW}$ and $dS_\text{SW}^\ast$ in the limit $\e_1,\e_2\to0$. They are expressed in terms of the two solutions of the SW curve written in \ref{sol_SW_curve}:
\begin{equation}
dS_\text{SW}=-\log(y_+(z))dz,\quad dS_\text{SW}^\ast=-\log(y_-(z))dz.
\end{equation}
It has been shown in the section \refOld{sec42} that they coincide withe limit of the vevs $\la dS_{\vec Y}\rag$ and $\la dS_{\vec Y}^\ast\rag$ respectively.

Before taking the limit of the qq-deformed relation, we would like to examine more closely the usual second Seiberg-Witten relation. The cycles $\CB_l$ consist of a path starting from infinity on the first sheet joining a point $\a_l$ on the $l$th branch cut, then going to the second sheet where it joins $\a_l$ to infinity again. If we identify the two paths, the contour integral can be written as an integral on the complex plane of the difference $dS^-_\text{SW}=dS_\text{SW}-dS_\text{SW}^\ast$:
\begin{equation}
\oint_{\CB_l}dS_\text{SW}=\int_{\infty}^{\a_l}dS_\text{SW}+\int_{\a_l}^\infty{dS_\text{SW}^\ast}=-\int_{\a_l}^\infty{dS^-_\text{SW}}.
\end{equation} 
The RHS does not depend on the choice of the point $\a_l$ on the branch cut. Thus, it is possible to integrate over the whole branch cut. Using the fact that the support of the profile density $f_l''(z)$ coincides with the $l$th branch cut, and that it is normalized to two \cite{Nekrasov2003a}, it is possible to write:
\begin{equation}\label{Wanted}
\oint_{\CB_l}dS_\text{SW}=-\hf\int_{\mathbb{R}}d\a f_l''(\a)\int_{\a}^\infty{dt\ \dfrac{dS^-_\text{SW}}{dt}},\quad\text{since}\quad \hf\int_{\mathbb{R}}f_l''(t)dt=1.
\end{equation} 
This expression will be compared to the second qqSW relation \ref{qqSW_II} in the limit $\e_1,\e_2\to0$.

The SW limit can be reached by taking the limit $\e_1\to0$ of the NS limit. In the NS limit, the second qSW relation has been obtained in \ref{limit_qqSW_I}. Introducing the density $\rho_u^{(l)}(t)$ of Bethe roots in the $l$th string, this relation takes the following form,
\begin{equation}
\dfrac{\p\CF_\text{NS}}{\p a_l}=\int_{\mathbb{R}}dt\rho_u^{(l)}(t)\int_{t}^\infty\dfrac{d^2S_\text{NS}^-(s)}{ds^2}ds,\quad\text{with}\quad \rho_u^{(l)}(t)=\e_1\sum_{i=1}^\infty\d(t-u_{(l,i)}).
\end{equation} 
In the NS limit, the Bethe roots are identified with the coordinates $\phi_x$ of boxes that can be added or removed from the Young diagram $\vec Y^\ast$ at the saddle point. It is then possible to use the formula \ref{def_profile} to relate the Young diagram profile with the density of Bethe roots,
\begin{equation}
f_l'(t)=-1+(1-e^{-\e_1\p_t})\sum_{i=1}^\infty\sign(t-u_{(l,i)}).
\end{equation} 
As a result, in the limit $\e_1\to0$, we can identify $f'_l(t)=-1+2\rho_u^{(l)}(t)$, and 
\begin{equation}\label{temp}
\dfrac{\p\CF_\text{SW}}{\p a_l}=\hf\int_{\mathbb{R}}dt(f'_l(t)+1)\int_{t}^\infty\dfrac{d^2S_\text{SW}^-}{ds^2}(s)\ ds.
\end{equation} 
This expression closely resemble \ref{Wanted}, but we still need to perform an integration by parts to show the agreement between the two. For this purpose, we need the following identity,
\begin{equation}
v_c(t)=\int_c^tds\int_{s}^\infty{\dfrac{d^2S_\text{SW}^-}{ds^{\prime2}}(s')\ ds'}=\int_t^\infty{\dfrac{dS_\text{SW}^-}{ds}(s)ds}-\int_c^\infty{\dfrac{dS_\text{SW}^-}{ds}(s)ds}
\end{equation} 
for any parameter $c\in\mathbb{R}$. The second term in the RHS is a simple integration constant, independent of $t$, and $v_c(t)=v_\infty(t)+\text{cste}$. Taking the integration by parts with respect to the variable $t$ in \ref{temp}, and using the previous identity, we find
\begin{equation}
\dfrac{\p\CF_\text{SW}}{\p a_l}=\hf\int_{\mathbb{R}}dt\left[\dfrac{d}{dt}\left((f'_l(t)+1)v_c(t)\right)-f_l''(t)v_c(t)\right].
\end{equation}
In this RHS, $v_c(t)$ can be replaced by $v_\infty(t)$ since the constant term does not contribute. Then, we notice that the boundary term is also vanishing, because $f'_l(t)$ tends to $-1$ as $t\to-\infty$ (see \cite{Nekrasov2003}), and $v_\infty(t)$ tends to zero as $t\to\infty$. As a result, it only remains
\begin{equation}
\dfrac{\p\CF_\text{SW}}{\p a_l}=-\hf\int_{\mathbb{R}}dt\ f_l''(t)\int_t^\infty{\dfrac{dS_\text{SW}^-}{ds}(s)ds}=\oint_{\CB_l}dS_\text{SW},
\end{equation}
according to \ref{Wanted}. Thus, we have reproduced the second SW relation from the limit $\e_1,\e_2\to0$ of the qqSW relation \ref{qqSW_II}.

\bibliographystyle{utphys}
\providecommand{\href}[2]{#2}\begingroup\raggedright\endgroup

\end{document}